\documentclass{aa}  

\usepackage{graphicx}
\usepackage{txfonts}
\usepackage{amssymb}
\usepackage{natbib}
\usepackage{rotating}
\usepackage{epsfig}
\usepackage{subfigure}
\usepackage{colortbl}
\usepackage{lscape}
\usepackage{tabularx}
\usepackage{longtable}
\usepackage{amsmath}

\newcommand{\nv}{N\,{\sc v}}

\newcommand{\siiv}{Si\,{\sc iv}}

\newcommand{\civ}{C\,{\sc iv}}

\newcommand{\aliii}{Al\,{\sc iii}}

\newcommand{\mgii}{Mg\,{\sc ii}}
\newcommand{\feii}{Fe\,{\sc ii}}
\newcommand{\feiii}{Fe\,{\sc iii}}
\newcommand{\hb}{H$\beta$}

\begin{document}

   \title{Investigating the radio-loud phase of broad absorption
line quasars}


\author{G. Bruni \inst{1,2}
\and J.I. Gonz\'alez-Serrano \inst{3}
\and M. Pedani \inst{4}
\and C.R. Benn \inst{5}
\and K.-H. Mack \inst{2}
\and J. Holt \inst{6}
\and \\F.M. Montenegro-Montes \inst{7}
\and F. Jiménez-Luj\'an \inst{8,9}}

   \institute{Max Planck Institute for Radio Astronomy, Auf dem H\"{u}gel, 69, D-53121  Bonn, Germany
   \and INAF-Istituto di Radioastronomia, via Piero Gobetti, 101, I-40129 Bologna, Italy
   \and Instituto de F\'isica de Cantabria (CSIC-Universidad de Cantabria), Avda. de los Castros s/n, E-39005 Santander, Spain
   \and INAF-Fundaci\'on Galileo Galilei, Rambla Jos\'e Ana Fern\'andez P\'erez, 7, 38712 Bre\~na Baja, TF, Spain
   \and Isaac Newton Group, Apartado 321, E-38700 Santa Cruz de La Palma, Spain
   \and Leiden Observatory, Leiden University, P.O. Box 9513, NL-2300 RA Leiden, The Netherlands
   \and European Southern Observatory, Alonso de C\'ordova 3107, Vitacura, Casilla 19001, Santiago, Chile
   \and Dpto. de F\'isica Moderna, Universidad de Cantabria, Avda de los Castros s/n, E-39005 Santander, Spain
   \and Dpto. de F\'isica, Universidad de Atacama, Copayapu, 485, Copiap\'o, Chile
   }

   \date{}

 
  \abstract
   {Broad absorption lines (BALs) are present in the spectra of $\sim$20\% of quasars (QSOs); this indicates fast outflows (up to 0.2$c$) that intercept the observer's line of sight. These
QSOs can be distinguished again into radio-loud (RL) BAL QSOs  and radio-quiet (RQ) BAL\ QSOs. The first are very rare, even four times less common than RQ BAL QSOs. The reason for this is still unclear and leaves open questions about the nature of the BAL-producing outflows and their connection with the radio jet.}
   {We explored the spectroscopic characteristics of RL and RQ BAL QSOs with the aim to find a possible explanation for the rarity of RL BAL QSOs.}
   {We identified two samples of genuine BAL QSOs from SDSS optical spectra, one RL and one RQ, in a suitable redshift interval (2.5$<z<$3.5) that allowed us to observe the \mgii~and \hb~emission lines in the adjacent near-infrared (NIR) band. We collected NIR spectra of the two samples using the Telescopio Nazionale Galileo (TNG, Canary Islands). By using relations known in the literature, we estimated the black-hole mass, the broad-line region radius, and the Eddington ratio of our objects and compared the two samples.}
   {We found no statistically significant differences from comparing the distributions of the cited physical quantities. This indicates
that they have similar geometries, accretion rates, and central black-hole masses, regardless of whether the radio-emitting jet is present or not.
}
{These results show that the central engine of BAL QSOs has the
same physical properties with and without a radio jet. The reasons for the rarity of RL BAL QSOs must reside in different environmental or evolutionary variables.}

   \keywords{quasars: absorption lines – galaxies: active – galaxies: evolution – radio continuum: galaxies}

   \maketitle
%
%

\section{Introduction}

Quasar outflows manifest most spectacularly as broad absorption lines (BALs) in the blue wings of prominent emission lines (e.g. \civ) in $\sim$20\% of optically selected quasars; they trace outflow velocities of up to  $\sim$0.2 c (\citealt{Hewett}). These absorption troughs present a complex structure, and in some cases ($\sim$20\%) can be detached by several thousand $km~s^{-1}$ from the corresponding emission line (\citealt{kor93}). The blue and red edges are often much narrower than the total width of the trough, and span hundreds of $km~s^{-1}$. Troughs are not saturated, indicating that the background nuclear regions are not completely covered.
Absorption by individual clouds cannot easily account for these observed features, while an opportunely oriented outflow could produce such effects. Different authors (\citealt{Murray}, \citealt{Elvis}) proposed that a fast outflow, emerging perpendicularly to the accretion disk, and then radially accelerated, could be at the origin of the BAL phenomenon. In some cases, an indication of the launching point of this assumed outflow originates in a partially absorbed Ly$\alpha$ emission line: this suggests that the outflow is outside the broad emission-line region (BLR), at a distance $>$0.1 pc from the quasar nucleus. BALs can also be variable and show a disappearance on a time scale of $\sim$100 years (\citealt{Filiz}), which suggests a possible reorientation of the outflow.
Most of BAL are associated with high-ionisation species, such as \civ, \siiv, \nv: quasars (QSOs) showing these features are then classified as HiBAL QSOs. A small fraction of these ($\sim$15\%) also show absorption from low-ionisation species such as \mgii ~or \aliii, and are known as LoBAL QSOs. An additional class is FeLoBAL QSOs, which are LoBAL QSOs that also show absorption produced by the \feii ~and \feiii ~lines. Absorbers similar to quasar BALs are seen in Seyfert 1 galaxies, albeit with lower outflow velocities, typically lower than a few hundred $km~s^{-1}$ (see contributions in \citealt{Crenshaw}).

No self-consistent physical model exists as yet for the acceleration of the outflowing gas in BAL quasars, or, if the filling factor is small (many small clouds), for its confinement. Possible mechanisms for the acceleration include radiation pressure, pressure from cosmic rays, or centrifugally driven magnetic disk winds (\citealt{deKool}). Radiation pressure is a popular candidate, but it is unclear how it can be sustained without over-ionising the gas. In this outflow scenario, the observed fraction of BAL QSOs over the total QSO population is explained as an orientation effect (\citealt{Weymann}, \citealt{Elvis}). An evolutionary scenario, proposed by different authors, suggests that BALs are present in an early stage of the QSO ($\sim$10–20\% of the total life), when the dust and gas cocoon is expelled, producing the absorption features (\citealt{Briggs}, \citealt{Sanders}).  

Before the advent of the FIRST Bright Quasar Survey (\citealt{Becker2}), only a few radio-loud BAL QSOs were identified and studied. Thanks to the better statistics offered by this catalogue, hints of an anticorrelation between the BAL phenomenon and radio-loudness were found: \cite{Becker2} found that the fraction of BAL QSOs with $\log R^{*}>2$ (radio-loudness $R^{*}=S_{5\rm{GHz}}/S_{2500~\AA}$, \citealt{Stocke}) is only $\sim$25\% of the total BAL QSOs population. \cite{Gregg}, in a work about Fanaroff-Riley II BAL QSOs, found an anticorrelation between radio-loudness and the BALs strength. He accounted for that by referring to the evolutionary scenario, in which the ejected cocoon would stifle the development of the radio-jet and lobes. \cite{Hewett}, studying the intrinsic fraction of BAL QSOs in optically selected samples, suggested that optically bright BAL quasars are half as likely as non-BALs to have $S_{1.4 \rm{GHz}}>1$ mJy. \cite{Becker1} first used the radio emission spectral index to derive the jet orientation of BAL QSOs, and noted that this can lie in wide range of possible angles.
After these works, different authors presented detailed studies about larger samples of radio-loud BAL QSOs, from which they reported a variety of possible orientations for the outflow, but also a variety of possible morphologies and ages for the radio source (\citealt{Montenegro, DiPompeo1, Bruni1, Bruni2}). This does not suggest a scenario that would be clearly referable to one of the proposed models.

\cite{Boroson} classified AGNs based on a principal-component analysis of AGN properties. This sorted the different observed types according to different combinations of $L/L_{Eddington}$ (luminosity as a fraction of Eddington luminosity) and $\dot M$ (the accretion rate). In this scheme, BAL QSOs are predicted to be objects with a high accretion rate, and radio-loud BAL QSOs with an even higher one. BAL outflows are crucial for understanding the physics of AGN because
1) they probe the inner regions of the accretion disk, and probably play a role in the accretion process by helping to shed angular momentum;
2) many BAL quasars are super-Eddington accretors, which offers a unique perspective on the changes in disk geometry (e.g. thickening) with accretion rate;
3) the highly energetic BAL outflows are probably related to other outflows seen in AGN (e.g. in radio galaxies) and their duty-cycle.

In the near-infrared (NIR) domain, it is possible to observe the UV-optical part of the QSO spectrum for objects with a redshift of between $\sim$2.5 and $\sim$3.5. In this portion of the spectrum, the BALs can be present at the blue-side of the \mgii~emission line. Stronger BAL features are associated with the \civ~emission line, which is detectable in the optical window. 
From the analysis of the (\mgii, \hb) emission lines and of the adjacent continuum emission, it is possible to study important characteristics of the QSO, from which the possible differences introduced by radio-loudness can be investigated.%
\begin{itemize}
        \item The FWHM of \mgii~and \hb~can be indicative of the central BH mass, according to the scaling relations given by \cite{Vestergaard} and \cite{Vestergaard2}.\\
        
        \item The Eddington ratio of the central BH can be estimated from the luminosity at 5100 \AA, as suggested by \cite{Kaspi}. \\

        \item The size of the broad-line region can be related to the luminosity at 5100 \AA~(\citealt{Kaspi, Kaspi05, Bentz}).

\end{itemize}

This paper is organized as follows: In Sect. 2 we present the sample selection, in Sect. 3 we describe the observations and data reduction, in Sect. 4 we discuss the data analysis and present the adopted method for estimating the derived quantities, and finally our results.

The cosmology adopted throughout the paper assumes a flat universe and
the following parameters: $H_{0}$=70 km s$^{-1}$ Mpc$^{-1}$,
$\Omega_{\Lambda}$=0.7, $\Omega_{M}$=0.3.

%

\begin{figure}
        \includegraphics[width=9cm]{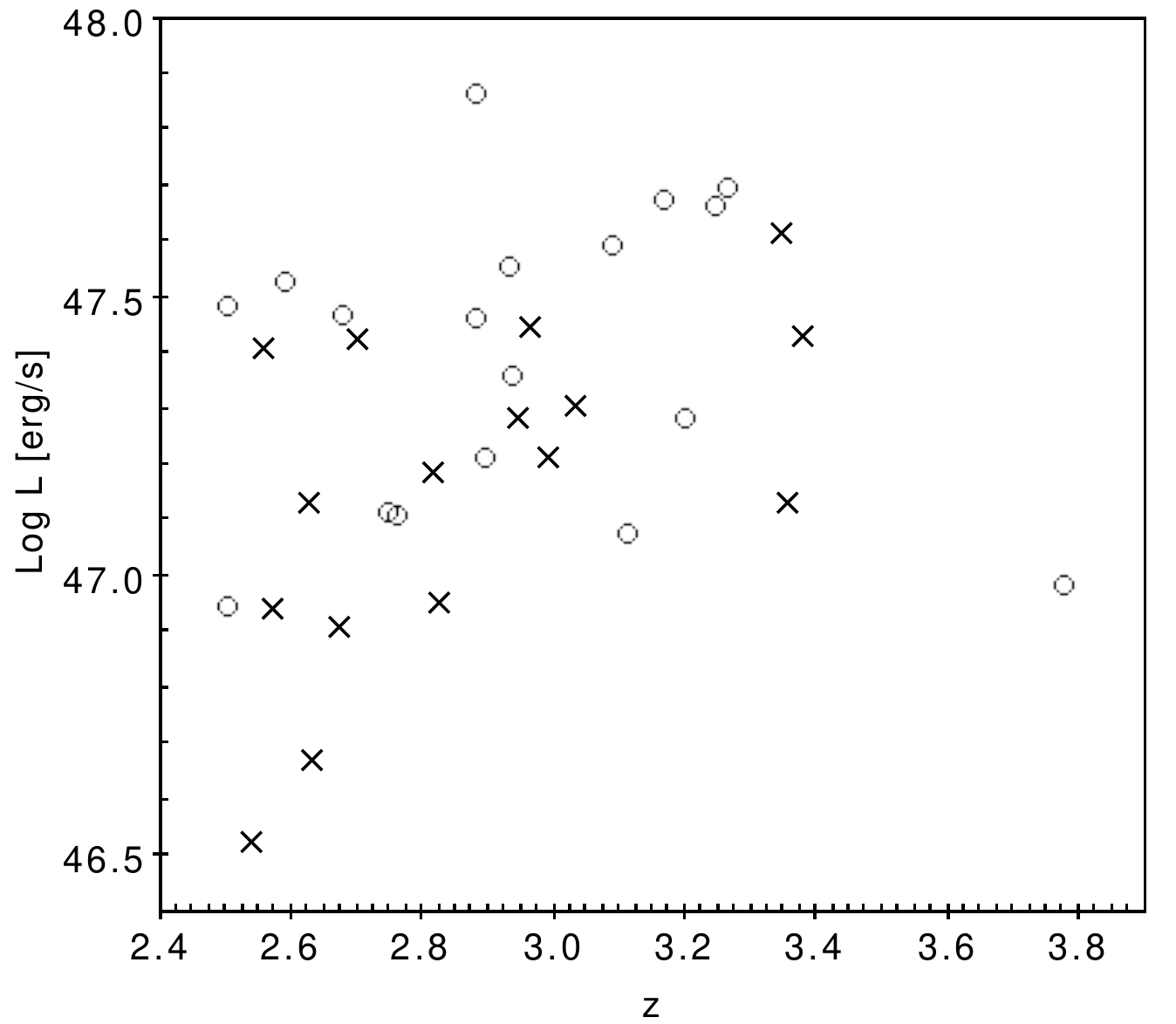}
        \caption{Bolometric luminosity (from \citealt{Shen}) \emph{vs} redshift for the RL (crosses) and RQ (circles) BAL QSOs of our optically bright sample.}
\label{sample}
\end{figure}

\begin{figure}
        \includegraphics[width=9cm]{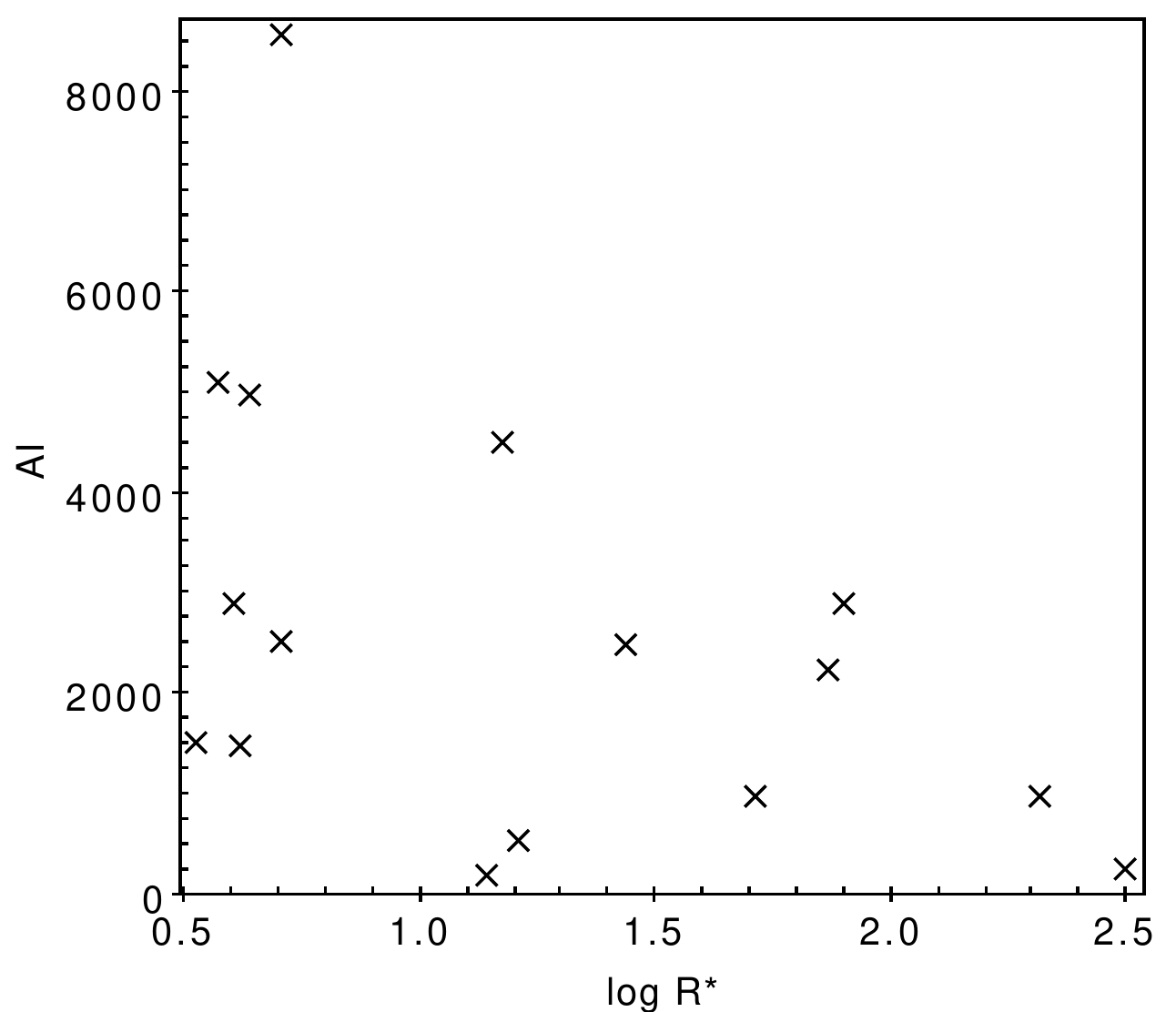}
        \caption{Absorption index \emph{vs} Log \emph{R*} for our optically bright sample. Values of \emph{R*} are taken from \cite{Shen}.}
\label{AI}
\end{figure}

%

\section{Optically bright sample}

The original samples of RL and RQ BAL QSOs studied in this work were built in 2007, starting from the fourth data release of the SDSS (DR4, \citealt{Adelman06}). As a first step, we selected all the QSOs from the SDSS DR4 with a redshift $z>$2.5: this threshold allowed us to visualise the \civ~emission line in the SDSS spectra so that we would recognise possible BAL features, and the \mgii~and \hb~emission lines in the NIR window. Then, we set an additional upper limit of r$<$19.5 to the apparent magnitude in r-band to obtain a better signal-to-noise ratio
(S/N) during observations. All spectra satisfying these criteria were visually inspected and classified as BAL or non-BAL QSOs: this yielded 209 BAL QSOs.

As a second step, this list of BAL QSOs was cross-correlated with the FIRST survey to divide
them into radio-loud (RL, $S_{1.4 \rm{GHz}}>1$ mJy) and radio-quiet (RQ, $S_{1.4 \rm{GHz}}=0$ mJy). From each of these two sub-samples, we selected the 21 RL BAL QSOs and the 23 RQ BAL QSOs with the brightest luminosity in r-band. 
The definition of radio-loudness considered in this work differs from that used by \citealt{Becker2}: since all of our objects have $S_{1.4 \rm{GHz}}>1$ mJy, we considered this condition sufficient to investigate the findings by \cite{Hewett}, i.e. the rarity of optically bright BAL QSOs among objects with $S_{1.4 \rm{GHz}}>1$ mJy. At a redshift of 2.5 (lower threshold of our sample) a detected flux density of 1 mJy in the FIRST survey corresponds to a luminosity $L_{1.4 \rm{GHz}}=10^{32.7}$ erg/s/Hz, which already exceeds the accepted threshold at which a source is considered radio-loud ($L_{1.4 \rm{GHz}}>10^{32.5}$ erg/s/Hz, \citealt{Gregg96}).

To refine the BAL classification and set it in line with our previous work, in 2013 we considered the objects' spectra from the latest version of the SDSS (DR9, \citealt{Ahn}) and performed a proper calculation of the absorption index (AI), as defined in \cite{Bruni1}. We used the classical definition by \cite{Hall}:
\begin{equation}
{\rm{AI}}=\int_{0}^{25000}(1-\frac{f(v)}{0.9})\cdot Cdv,
\end{equation}
but with the change that the parameter $C$ is unity over contiguous troughs of at least 1000 km s$^{-1}$ (as in \citealt{Trump}), 
and we considered as genuine BAL QSOs only objects with an AI$>$100. To perform this calculation, we integrated the
spectral region between the peaks of the \civ~and \siiv~emission lines to up to 25000 km s$^{-1}$ from the former.
As a result, five RL and RQ objects were rejected. 

Finally, thanks to the the joint analysis of the optical and infrared spectra collected for this work (see next section), we were also able to classify the objects into HiBAL, LoBAL, and FeLoBAL QSOs. The final list of 16 RL and 18 RQ BAL QSOs studied here, together with the calculated AI and classification, is presented in Table \ref{sources}. In Fig \ref{sample}, the distribution of bolometric luminosity \emph{vs} redshift is given for the two samples. 

In Fig. \ref{AI} we plot the AI \emph{vs} Log \emph{R*} for the 16 RL objects: we find an anticorrelation analogous 
to that found by \cite{Gregg}.

\begin{table*}
 \centering
  \caption{Samples of 16 RL (top) and 18 RQ (bottom) BAL QSOs studied in this work. Col. 1: SDSS ID; Col. 2: source ID in this paper; Col. 3,4: optical coordinates in J2000.0 from SDSS; Col. 5: redshift as measured from the SDSS DR4 - the asterisked value was corrected by us by mean of a \civ~line identification; Col. 6,7: magnitudes in r and k bands from SDSS DR4 and 2MASS Point Source Catalogue (\citealt{2mass}), respectively; Col. 8: peak flux density at 1.4 GHz from the FIRST; Col. 9,10: absorption index and BAL type from this work.}
  \label{sources}

\scalebox{0.95}{
  \begin{tabular}{ccccccccccc}
  \hline
   \multicolumn{1}{c}{SDSS ID}         &
   \multicolumn{1}{c}{ID}         &
   \multicolumn{1}{c}{RA}           & 
   \multicolumn{1}{c}{DEC}          & 
   \multicolumn{1}{c}{z}            & 
   \multicolumn{1}{c}{r}            & 
   \multicolumn{1}{c}{K}            & 
   \multicolumn{1}{c}{FIRST}        &
   \multicolumn{1}{c}{AI}        &
   \multicolumn{1}{c}{Type}        \\
      
   \multicolumn{1}{c}{}                 &
   \multicolumn{1}{c}{}                 &
   \multicolumn{1}{c}{(J2000)}      & 
   \multicolumn{1}{c}{(J2000)}      &
   \multicolumn{1}{c}{}             &
   \multicolumn{1}{c}{(mag)}             &
   \multicolumn{1}{c}{(mag)}             &
   \multicolumn{1}{c}{(mJy)}        &
   \multicolumn{1}{c}{(km/s)}             &   
   \multicolumn{1}{c}{}             \\
    
   \multicolumn{1}{c}{(1)}          &
   \multicolumn{1}{c}{(2)}          &
   \multicolumn{1}{c}{(3)}              &
   \multicolumn{1}{c}{(4)}          &
   \multicolumn{1}{c}{(5)}          &
   \multicolumn{1}{c}{(6)}          &
   \multicolumn{1}{c}{(7)}          &
   \multicolumn{1}{c}{(8)}          &
   \multicolumn{1}{c}{(9)}          &
   \multicolumn{1}{c}{(10)}        \\
  
\hline
\object{SDSS J084401.95+050357.8} & 0844+05            & 08 44 01.96 &  +05 03 57.9   & 3.34           &  18.19 &14.19         &   7.2         & 5104          &       Lo      \\
\object{SDSS J111038.46+421715.6} &1110+42             & 11 10 38.46 &  +42 17 15.7   & ~~2.52*        &  18.70 &14.56         &   2.8         & 2879          &       Lo\\    
\object{SDSS J115205.40+042952.9} &1152+04             & 11 52 05.41 &  +04 29 52.9   & 3.35           &  19.08 &-                     &   1.9   & 1489          &       Hi      \\
\object{SDSS J115947.10+413659.1} &1159+41              & 11 59 47.10 &  +41 36 59.1   & 2.94           &  18.97 &-                     &   2.5   & ~~526 &       Hi      \\
\object{SDSS J121420.09+514924.7} &1214+51              & 12 14 20.10 &  +51 49 24.8   & 2.63           &  18.88 &15.94         &   3.7         & 2495          &       FeLo\\  
\object{SDSS J123651.28+453334.5} &1236+45              & 12 36 51.28 &  +45 33 34.5   & 2.56           &  18.11 &14.97         &   1.9         & 1453          &       Hi      \\
\object{SDSS J130756.73+042215.5} &1307+04              & 13 07 56.73 &  +04 22 15.5   & 3.02           &  18.39 &15.10         &  14.3~~         & ~~188 &       Hi      \\
\object{SDSS J132703.21+031311.2} &1327+03              & 13 27 03.22 &  +03 13 11.3   & 2.82           &  19.26 &15.06         &  60.7~~         & ~~240 &       Hi      \\
\object{SDSS J135909.92+470826.5} &1359+47              & 13 59 09.93 &  +47 08 26.5   & 2.53           &  19.09 &-                     &   2.2   & 4950          &       Lo      \\
\object{SDSS J141334.38+421201.7} &1413+42              & 14 13 34.38 &  +42 12 01.7   & 2.81           &  18.98 &15.54         &  17.8~~       & 2236            &       Hi      \\
\object{SDSS J142610.59+441124.0} &1426+44              & 14 26 10.59 &  +44 11 24.0   & 2.67           &  18.98 &-                     &   6.8   & 4485          &       FeLo    \\
\object{SDSS J145910.13+425213.2} &1459+42              & 14 59 10.14 &  +42 52 13.2   & 2.96           &  18.40 &15.84         &  12.8~~         & ~~972 &       Hi      \\
\object{SDSS J151601.52+430931.4} &1516+43              & 15 16 01.51 &  +43 09 31.5   & 2.63           &  18.79 &15.24         &   1.3         & 8564          &       Lo      \\
\object{SDSS J162453.47+375806.6} &1624+37              & 16 24 53.48 &  +37 58 06.6   & 3.38           &  18.74 &15.57         &  56.1~~         & ~~976 &       Hi      \\
\object{SDSS J162559.89+485817.4} &1625+48              & 16 25 59.90 &  +48 58 17.5   & 2.72           &  18.31 &15.06         &  25.3~~         & 2869          &       Lo      \\
\object{SDSS J163750.36+322313.8} &1637+32              & 16 37 50.37 &  +32 23 13.8   & 2.99           &  19.15 &-                     &   5.1   & 2474          &       FeLo    \\
&&&&&&&&&\\
\hline
&&&&&&&&&\\
\object{SDSS J031609.83+004043.0} &0316+00       &  03 16 09.84 &  +00 40 43.2 &  2.92&  18.98 &    -          &-& ~~397       & Hi            \\
\object{SDSS J073535.44+374450.4} &0735+37       &  07 35 35.45 &  +37 44 50.4 &  2.75&  18.89 &    -          &-& 5549                & Hi              \\
\object{SDSS J074628.70+301419.0} &0746+30       &  07 46 28.72 &  +30 14 19.0 &  3.11&  18.81 &    15.15      &-& 8034                & Lo              \\
\object{SDSS J080006.32+443555.6} &0800+44       &  08 00 06.32 &  +44 35 55.6 &  2.50&  18.95 &    15.27      &-& 4208                & Hi              \\
\object{SDSS J091307.82+442014.3} &0913+44       &  09 13 07.83 &  +44 20 14.4 &  2.93&  18.34 &    14.89      &-& 4832                & FeLo            \\
\object{SDSS J095858.14+362318.9} &0958+36       &  09 58 58.14 &  +36 23 18.9 &  2.67&  18.15 &    14.64      &-& 5290                & Lo              \\
\object{SDSS J102009.99+104002.7} &1020+10       &  10 20 09.99 &  +10 40 02.8 &  3.16&  18.14 &    15.32      &-& ~~536       & Hi            \\
\object{SDSS J105904.68+121024.0} &1059+12       &  10 59 04.69 &  +12 10 24.0 &  2.50&  17.85 &    15.04       &-& ~~888      & Hi            \\
\object{SDSS J131912.39+534720.5} &1319+53       &  13 19 12.41 &  +53 47 20.6 &  3.09&  18.26 &    15.29       &-& 2521               & Hi              \\
\object{SDSS J134722.83+465428.5} &1347+46       &  13 47 22.84 &  +46 54 28.6 &  2.93&  18.30  &    15.16      &-& 3766               & FeLo            \\
\object{SDSS J140006.88+412142.5} &1400+41       &  14 00 06.89 &  +41 21 42.6 &  2.76&  18.39 &    15.45       &-& 3730               & Lo              \\
\object{SDSS J140745.50+403702.2} &1407+40       &  14 07 45.50 &  +40 37 02.3 &  3.20 &  18.82 &    14.63              &-& 6589               & Lo              \\
\object{SDSS J142543.32+540619.3} &1425+54       &  14 25 43.32 &  +54 06 19.4 &  3.24&  18.04 &    15.24       &-& ~~269      & Hi            \\
\object{SDSS J150332.17+364118.0} &1503+36       &  15 03 32.18 &  +36 41 18.0 &  3.26&  18.22 &    15.01       &-& 2061               & FeLo            \\
\object{SDSS J152553.89+513649.1} &1525+51       &  15 25 53.90 &  +51 36 49.3 &  2.88&  17.19 &    14.02       &-& 1269               & FeLo            \\
\object{SDSS J153715.74+582933.9} &1537+58       &  15 37 15.74 &  +58 29 33.8 &  2.59&  17.61 &    14.71       &-& 3206               & FeLo            \\
\object{SDSS J164151.83+385434.2} &1641+38       &  16 41 51.84 &  +38 54 34.2 &  3.77&  18.88 &    15.33       &-& 5079               & Hi              \\
\object{SDSS J164219.88+445124.0} &1642+44       &  16 42 19.89 &  +44 51 24.0 &  2.88&  18.24 &    14.91       &-& 2787               & Lo              \\
\hline
\end{tabular}}
\end{table*}
%
%
%
%
%
\section{NIR observations and data reduction}

These observations were obtained with the Near Infrared Camera Spectrograph (NICS) at the 3.58m Telescopio Nazionale Galileo (TNG). NICS (\citealt{Baffa}) offers a unique, high-sensitivity and low-resolution observing mode that uses an Amici Prism as a dispersing element (\citealt{Oliva}). In this mode it is possible to obtain a spectrum from 0.8 to 2.4 microns in a single exposure with a very high throughput.
Spectral resolution with 1" slit (the default setup) is R$\sim$50 or $\sim$6000 $km/s$.
This observing mode is appropriate to study typical BAL QSOs with broad ($>$4000 $km/s$) emission lines and absorption systems.

Observations were carried out in five observing runs during the 2008 ITP (International Time La Palma): Jan 25, Mar 14-15, and Jun 25-26. A total of 44 QSOs were observed with a non-repetitive ABBA nodding pattern along the slit; 
typical integration time was 24 minutes per target. 

Data reduction was performed with a semi-automatic pipeline, taking as input the QSO spectrum, that of a telluric standard star (FV-GV) taken at similar airmass (to remove the telluric absorption features), and a look-up table for the wavelength calibration. 
The atmospheric transparency is a free parameter that is changed manually to take into account any variation of the sky transparency along the night. This allows the atmospheric residuals in the final spectrum to be minimized. Small shifts in the wavelength calibration solution can also be introduced at this point of the process to perfectly
match the broad telluric absorption features of the standard and the QSO.
Absolute flux calibration was obtained by scaling the spectrophotometric flux of the QSO to that of the telluric standard at H band. To improve this calibration, we renormalised the spectra by comparing
them with the flux density of the SDSS at the overlapping wavelength of 9000 \AA. The obtained scaling factors vary in a range from $\sim$0.3 to $\sim$3. We considered an uncertainty of 10\% on the absolute flux density calibration, and thus on the derived objects luminosities presented in Table \ref{results1}. 
The spectra of the RL and RQ BAL QSOs studied here are shown in Fig. \ref{spectra}.
%
%
%
%
%
%
\section{Results and discussion}

We analysed the spectra using the SPLAT\footnote{http://star-www.dur.ac.uk/\textasciitilde pdraper/splat/splat-vo/splat-vo.html} package from the European Virtual Observatory. We fitted the continuum using a second- or third-grade polynomial function, and the emission lines using a Gaussian with free parameters (centre, peak, sigma). Standard UV and optical-line identifiers were used. 

Our main goal was to extract the FWHM of the \mgii~ and \hb~ emission lines to estimate the central black-hole masses of the two samples. In the wavelength range from 0.8 to 2.4 microns there are two critical intervals because of very low atmospheric transmittance: 1.35 - 1.44 microns and 1.81 - 1.94 microns. When one of the lines fell in these ranges it was not possible to perform the fit, because of the superposition with atmospheric features. We calculated the intrinsic FWHM as 
\begin{equation}
\rm{FWHM}_{I}=\sqrt{\rm{FWHM}^{2}_{obs}-\rm{FWHM}^{2}_{res}}, 
\end{equation}
where $\rm{FWHM}_{obs}$ is the value obtained from the Gaussian fit, and $\rm{FWHM}_{res}$ is the wavelength uncertainty given by the spectral resolution for the corresponding emission line. The obtained FWHMs are presented in Table \ref{results1} together with the continuum luminosity at 3000 and 5100 \AA. \mgii~ can present a double-peaked emission: in this case, we performed a double-Gaussian profile fit and combined the two FWHM by the quadratic sum. 

In the following, we show the adopted methods and the results obtained following the various relations in the literature. The derived quantities are compared between the RL and RQ samples by mean of the Student's t-test, assuming different variances for the distributions (Welch t-test). This test allows us to compare the mean values of the studied quantities, while a point-to-point comparison might be misleading, given the quite large uncertainties of the resulting estimates.


\subsection{Black-hole mass estimation}

As a first step, we  inferred the different properties of the two groups of BAL QSOs by estimating the central black-hole (BH) mass. If they were an intrinsically different group of objects, different physical properties would be present, and the central black-hole mass can be an important variable for the dynamics of the central environment of the AGN.

In the literature, various relations are present to infer the mass of the central BH from single-epoch spectra: these are empirical relations, derived from reverberation mapping (RM, see \citealt{Peterson11} for a review), related to the FWHM of the strong UV emission lines in the optical-UV domain (\civ, \mgii, \hb) and with the continuum luminosity at a given wavelength. These relations are applicable with the caveats that no absorption is present in the line profile, and a good S/N  is available from measurements, otherwise the uncertain FWHM measurement can introduce errors or even systematic biases in the final estimate. For this reason, we decided to discard the \civ~line from this analysis, since it presents the BAL feature.

Another important condition for performing a good estimate through scaling relations is that the chosen formulae are on the same mass-scale: at present, the only sets of relations for \hb~and \mgii~with this property are those presented by \cite{Vestergaard} and \cite{Vestergaard2}. Thus, the two relations we used for our estimate are the following:

\begin{equation}
M_{\rm BH}~[M_\odot]= 10^{6.86}\left[ \frac{\rm FWHM(Mg II)}{1000~km~s^{-1}} \right]^2\left[\frac{\lambda \it L_{\lambda} {\rm (3000\,\AA)}}{10^{44} \rm erg~s^{-1}}\right]^{0.50} 
\label{eq:mgii}
\end{equation}
from \cite{Vestergaard2}, where $L_{\lambda} {\rm (3000\,\AA)}$ is the rest-frame luminosity at 3000\AA.
This relation is given with a 1-$\sigma$ scatter of 0.55 dex.

The second estimator we used has been derived by \cite{Vestergaard}, and is related with the \hb~ line:
\begin{equation}
M_{\rm BH}~[M_\odot] =10^{6.91}\left[ \frac{\rm FWHM(H\beta)}{1000~km~s^{-1}} \right]^2\left[\frac{\lambda \it L_{\lambda} {\rm (5100\,\AA)}}{10^{44} \rm erg~s^{-1}}\right]^{0.50} 
\label{eq:hb}
,\end{equation}                                                   
where $L_{\lambda} {\rm (5100\,\AA)}$ is the rest-frame luminosity at 5100 \AA.
This relation is given with a 1-$\sigma$ scatter of $\pm$0.43 dex.

These relations take advantage of the most recent updated analysis of the reverberation-mapping sample (\citealt{Peterson04}) and of the improved radius-luminosity (R-L) relations between the BLR size and continuum luminosity  (\citealt{Kaspi05,Bentz}).
Results are presented in Table \ref{results1} together with the FWHM of the lines as extracted from the spectra and the continuum luminosity.  We applied the K-correction both to the continuum and the FWHM.

\begin{table*}
 \centering
  \caption{Measured (rest-frame) and derived quantities for the sample of 16 RL (top) and 18 RQ (bottom) BAL QSOs. Column 6 reports the bolometric luminosity from \cite{Shen}.}
  \label{results1}
\renewcommand\tabcolsep{4pt}
\scalebox{0.95}{
  \begin{tabular}{cccccccccc}
  \hline
   \multicolumn{1}{c}{ID}                               &
   \multicolumn{1}{c}{FWHM \mgii}                       & 
   \multicolumn{1}{c}{FWHM \hb}                         & 
   \multicolumn{1}{c}{log($\lambda$L$_{3000}$)}                 & 
   \multicolumn{1}{c}{log($\lambda$L$_{5100}$)}                 &
   \multicolumn{1}{c}{log(L$_{Bol}$)}           &
   \multicolumn{1}{c}{log(M$_{BH}^{MgII}$)}     &
   \multicolumn{1}{c}{log(M$_{BH}^{H\beta}$)}   &            
   \multicolumn{1}{c}{BLR radius}                       &
   \multicolumn{1}{c}{L$_{Bol}$/L$_{Edd}$}              \\

   \multicolumn{1}{c}{}                                 &
   \multicolumn{1}{c}{(km/s)}                   & 
   \multicolumn{1}{c}{(km/s)}                   &
   \multicolumn{1}{c}{(erg/s)}      &
   \multicolumn{1}{c}{(erg/s)}      &   
   \multicolumn{1}{c}{(erg/s)}      &
   \multicolumn{1}{c}{(M$_\odot$)}     &
   \multicolumn{1}{c}{(M$_\odot$)}     &
   \multicolumn{1}{c}{(light-days)}             &
   \multicolumn{1}{c}{}                         \\

   \multicolumn{1}{c}{(1)}          &
   \multicolumn{1}{c}{(2)}          &
   \multicolumn{1}{c}{(3)}          &
   \multicolumn{1}{c}{(4)}          &
   \multicolumn{1}{c}{(5)}          &
   \multicolumn{1}{c}{(6)}          &  
   \multicolumn{1}{c}{(7)}          &
   \multicolumn{1}{c}{(8)}          &   
   \multicolumn{1}{c}{(9)}          &
   \multicolumn{1}{c}{(10)}          \\

\hline
0844+05 &       9110$\pm$1380           &       3610$\pm$1380   &       47.42              &     47.35   & 47.61  &      10.49$\pm$0.13~~        &       9.70$\pm$0.33   &              1060$\pm$50~~     &       0.11$\pm$0.03   \\
1110+42 &      13770$\pm$1700~~         &       -               &       46.79             &      46.34   & 46.94  &      10.53$\pm$0.11~~        &       -                 &               330$\pm$17      &       -       \\
1152+04 &       5350$\pm$1380           &       -               &       46.55             &      46.28   & 47.13  &       9.59$\pm$0.22          &       -                 &               309$\pm$15      &       -       \\
1159+41 &       5450$\pm$1520           &       -               &       46.59             &      46.19   & 47.28  &       9.63$\pm$0.24          &       -                 &               279$\pm$14      &       -       \\
1214+51 &       7780$\pm$1650           &       4150$\pm$1650   &       46.78             &      46.55   & 46.67  &      10.03$\pm$0.19~~        &       9.42$\pm$0.35   &               419$\pm$21      &       0.03$\pm$0.01   \\
1236+45 &     12970$\pm$1680~~          &       6667$\pm$1680   &       46.98             &      46.72   & 47.40  &      10.57$\pm$0.11~~        &       9.92$\pm$0.22   &               513$\pm$26      &       0.05$\pm$0.01   \\
1307+04 &       7990$\pm$1490           &       -               &       47.00             &      46.69   & 47.30  &      10.16$\pm$0.16~~        &       -                 &               491$\pm$25      &       0.11$\pm$0.04   \\
1327+03 &       9860$\pm$1570           &       -               &       46.54             &      46.19   & 46.95  &      10.12$\pm$0.14~~        &       -                 &               279$\pm$14      &       0.05$\pm$0.02   \\
1359+47 &       4250$\pm$1690           &       -               &       46.50             &      46.23   & 46.52  &       9.37$\pm$0.35          &       -                 &               291$\pm$15      &       -       \\
1413+42 &       5960$\pm$1570           &       -               &       46.72             &      46.45   & 47.18  &       9.77$\pm$0.23          &       -                 &               374$\pm$19      &       -       \\
1426+44 &       -                       &       7030$\pm$1630   &       46.69             &      46.62   & 46.91  &      -               &       9.91$\pm$0.20         &               455$\pm$23      &       -   \\
1459+42 &       7520$\pm$1510           &       -               &       46.60             &      46.05   & 47.44  &       9.91$\pm$0.18          &       -                 &               236$\pm$12      &       0.27$\pm$0.11   \\
1516+43 &       5830$\pm$1650           &       4340$\pm$1650   &       46.77             &      46.56   & 47.13  &       9.77$\pm$0.25          &       9.47$\pm$0.33   &               426$\pm$21      &       -         \\
1624+37 &    10850$\pm$1370~~           &       -               &       46.85             &      46.68   & 47.42  &      10.36$\pm$0.11~~        &       -                 &               486$\pm$24      &       0.09$\pm$0.02   \\
1625+48 &    15080$\pm$1610~~           &       -               &    46.91             &     46.67   & 47.42  &      10.67$\pm$0.10~~        &       -                 &               484$\pm$24      &       0.04$\pm$0.01   \\
1637+32 &       -                       &       -               &       46.65             &      46.52   & 47.21  &       -              &       -                &               405$\pm$20      &       -               \\
&&&&&&&&&\\
\hline
&&&&&&&&&\\
0316+00 &       -               &       -               &       46.54   &       46.18   &  47.21  &        -                      &       -                 &  274$\pm$14      &        -              \\
0735+37 &       3510$\pm$1600   &       -               &       46.58   &       46.24   &  47.11  &        9.24$\pm$0.40          &       -                 &  294$\pm$15      &        -      \\
0746+30 &       -               &       -               &       47.01   &       46.77   &  47.07  &        -                      &       -                 &  543$\pm$27      &        -              \\
0800+44 &       5030$\pm$1710   &       8910$\pm$1710   &       46.62   &       46.37   &  46.95  &        9.57$\pm$0.30          &      10.00$\pm$0.17~~   &     342$\pm$17      &        0.07$\pm$0.03  \\
0913+44 &       6730$\pm$1520   &       1170$\pm$1520   &       47.06   &       46.90   &  47.36  &       10.04$\pm$0.20~~        &       -     & 630$\pm$31      &        0.17$\pm$0.07   \\
0958+36 &       -               &       -               &       46.88   &       46.57   &  47.47  &        -                      &       -                 &  432$\pm$22      &        -              \\
1020+10 &       -               &       -               &       46.90   &       46.74   &  47.67  &        -                      &       -                 &  522$\pm$26      &        -              \\
1059+12 &       9090$\pm$1710   &       8750$\pm$1710   &       46.95   &       46.63   &  47.48  &       10.25$\pm$0.17~~        &      10.11$\pm$0.17~~   &     460$\pm$23      &        0.14$\pm$0.05  \\
1319+53 &       9090$\pm$1470   &       -               &       46.74   &       46.34   &  47.59  &       10.15$\pm$0.14~~        &       -                 &  331$\pm$17      &        0.22$\pm$0.07  \\
1347+46 &       3120$\pm$1520   &       6380$\pm$1520   &       46.99   &       46.74   &  47.55  &        9.34$\pm$0.42          &       9.89$\pm$0.21     &   524$\pm$26      &        0.36$\pm$0.17  \\
1400+41 &       -               &       -               &       46.88   &       46.55   &  47.10  &        -                      &       -                 &  420$\pm$21      &        -              \\
1407+40 &       9270$\pm$1430   &       6720$\pm$1430   &       47.12   &       46.96   &  47.28  &       10.35$\pm$0.14~~        &      10.05$\pm$0.19~~   &     674$\pm$34      &        0.07$\pm$0.02  \\
1425+54 &      12080$\pm$1410~~ &       -               &       46.93   &       46.50   &  47.66  &       10.49$\pm$0.10~~        &       -                 &  397$\pm$20      &        0.12$\pm$0.03  \\
1503+36 &       4610$\pm$1410   &       5070$\pm$1410   &       47.13   &       46.81   &  47.69  &        9.75$\pm$0.27          &       9.73$\pm$0.24     &   568$\pm$28      &        -      \\
1525+51 &       8140$\pm$1540   &       -               &       47.41   &       47.22   &  47.86  &       10.39$\pm$0.17~~        &       -                 &  909$\pm$45      &        0.24$\pm$0.09  \\
1537+58 &       1360$\pm$1670   &       7740$\pm$1670   &       47.06   &       46.79   &  47.52  &        -      &      10.08$\pm$0.19~~   &     552$\pm$28      &        0.22$\pm$0.10   \\
1641+38 &       7280$\pm$1250   &       -               &       47.01   &       46.87   &  46.98  &       10.09$\pm$0.15~~        &        -                &  608$\pm$30      &        0.06$\pm$0.02  \\
1642+44 &       9200$\pm$1540   &       -               &       47.05   &       46.77   &  47.46  &       10.31$\pm$0.15~~        &        -                &  544$\pm$27      &        0.11$\pm$0.04  \\
\hline
\end{tabular}
} 
\end{table*}

The obtained masses can span more than one order of magnitudes for both samples.
In some cases, the estimate from \mgii~ could not be performed because of the strong absorption of the line. Estimates from 
the \hb~line were possible only when the line had a sufficiently S/N to allow a Gaussian fit. 
The estimates from the \mgii~line gives average values of 10.07 dex $M_{\odot}$ ($\sigma=0.41$ dex) for RL BAL\ QSOs and 10.00 dex $M_{\odot}$ ($\sigma=0.42$ dex) for RQ BAL QSOs. Estimates from the \hb~emission line were possible only for a few objects (five RL \emph{vs} 6 RQ), and present an average value of 9.68 dex $M_{\odot}$ ($\sigma=0.24$ dex) for RL and 9.98 dex $M_{\odot}$ ($\sigma=0.14$ dex) for RQ objects. 
Comparing the two distributions from the \mgii~estimator with the Student's  t-test (ST), we obtained a probability of 33\% for the two distributions to be different, which is far below the conventional 95\% threshold. Thus, we do not see any significant difference for the BH mass distributions of RL and RQ BAL QSOs.


\subsection{Eddington ratio}

From the continuum luminosity we can derive another quantity useful to test the differences between the two groups of BAL QSOs: the Eddington ratio. \cite{Boroson} explained the observable spectral characteristics of QSOs as driven from two principal quantities: the Eddington ratio ($L_{Bol}/L_{Edd}$) and the accretion rate ($\dot{M}$). In their diagram BAL QSOs occupy a corner because
they are peculiar objects with both high $\dot{M}$ and $L_{Bol}/L_{Edd}$. 

To perform the calculation, we used the bolometric luminosity from the quasar catalogue of \cite{Shen} and our mass estimate from the \mgii~line. Only estimates with an error $<$50\% of the value are reported. In three cases (0800+44, 1347+46, 1537+58) we used the estimate from the \hb~line, since the former had an error $>$50\% of the value. Results are presented in Table \ref{results1}. No objects from the RL or RQ samples show a super-Eddington luminosity. Mean values are 0.09 ($\sigma=0.08$) for RL and 0.16 ($\sigma=0.09$) for RQ samples. An ST test gives a probability of 90\% for the two distributions to be different, which still not suggests a significant intrinsic difference in the accretion rate for the two groups of objects.


\subsection{Broad-line region radius}

\cite{Vestergaard} derived relation (\ref{eq:hb}) starting from the broad-line region (BLR) radius,
obtained from RM, and from the velocity of hydrogen clouds, approximated by the FWHM of the 
\hb~emission line. It has been found that the BLR radius is correlated with the continuum luminosity 
(\citealt{Kaspi,Kaspi05}). \cite{Bentz} updated these results taking into account the contribution of the
host galaxy starlight. This last version is consistent with the method of \cite{Vestergaard}, and is
\begin{equation}
R_{BLR}=A\cdot\left[\frac{\lambda L_{\lambda}(5100)}{10^{44}\rm{erg~s^{-1}}}\right]^{0.5}~\rm{lt-days}
,\end{equation}
where the coefficient $A$ is the scaling factor. The main difference with respect to the 
formula proposed by \cite{Kaspi05} is the exponent of the luminosity, given as 0.69$\pm$0.5 in that work, which corresponds to the slope of the correlation in logarithmic scale. The value of the
constant $A$, as the value of the slope, depends on the method used to interpolate data (BCES or FITEXY, see \citealt{Bentz} for further details): both authors used the two methods and compared the values obtained for the scaling factor. In this work we decided to use the mean value proposed by \cite{Kaspi05} for the scaling factor (A=22.3$\pm$2.1) and applied the correction provided by \cite{Bentz} for the slope.

We obtained a mean BLR radius of 427 light-days for RL ($\sigma$=191) and 501 light-days for RQ BAL QSOs ($\sigma$=155). A Student's t-Test results in a probability of 77\% for the two distributions to be different, which again does not suggest different geometries for the two samples.


\subsection{Previous works in the literature}
 
In the past years, the interest of the community in the BAL phenomenon has increased thanks to the better capabilities
offered by new-generation instrumentation. In particular, during the elaboration of our work, three infrared studies of radio-loud BAL QSOs have been published. \cite{DiPompeo2} found a mid- to near-infrared excess in a sample of 72 RL BAL QSOs, 
with respect to unabsorbed QSOs. The authors suggested that this effect might be due to the evolutionary stage of the objects, but since their previous study of the same sample in the radio band suggested a mild orientation, a merging of the evolutionary and orientation scenario is considered to explain the phenomenon. \cite{Runnoe} presented a spectroscopic study of eight RL BAL QSOs and found no significant differences between RL and RQ BAL QSOs, but dissimilar properties between RL BAL and RL non-BAL QSOs. Eddington ratios have been calculated in that work, which yielded only one super-Eddington RL BAL QSO with a ratio higher than 8 (12.5\%). Finally, a Herschel-ATLAS study of far-infrared properties of BAL QSOs (\citealt{Cao}) showed that there are no differences in terms of star formation rate between Hi-BAL and non-BAL QSOs, which excludes the
possibility that the BAL class is in a particular young, star-formation phase. 

These results confirm the controversial and heterogeneous observational properties found in previous works about BAL QSOs and provide
no clear evidence about the nature of these objects.


\section{Conclusions}

We investigated the spectral properties of RL \emph{vs} RQ QSOs in the NIR domain by measuring the main emission lines and continuum to estimate physical quantities related to the central kpc of these objects. Our sample is composed only of optically bright BAL QSOs (r$<$19.5), which needs to be considered in a comparison with other samples. 

The results can be summarised as follows:

\begin{itemize}

\item the central BH masses do not show significant differences between RL and RQ BAL QSOs. 
This excludes the scenario in which possible differences  due
to the BH mass are at the origin of the rarity of RL BAL QSOs.\\ 

\item The Eddington ratio distributions derived for our samples of RL and RQ BAL QSOs are similar. This suggests similar accretion rates for the two groups of objects.\\

\item We did not find any significant difference between the mean BLR radius of RL and RQ BAL QSOs. This implies a similar geometry and dynamic for the central region of RL and RQ BAL QSOs.
\end{itemize}

In a future work, we will investigate the optical spectroscopic properties of the same samples with high-resolution spectra collected at the William Herschel Telescope (WHT, Canary Islands).

\begin{acknowledgements}

The authors acknowledge financial support from the Spanish Ministerio de Ciencia e Innovaci\'on under projects AYA2008-06311-C02-02 and AYA2011-29517-C03-02. Part of this work was supported by the COST Action MP0905 "Black Holes in a Violent Universe”. Based on observations made with the Italian Telescopio Nazionale Galileo (TNG) operated on the island of La Palma by the Fundaci\'on Galileo Galilei of the INAF (Istituto Nazionale di Astrofisica) at the Spanish Observatorio del Roque de los Muchachos of the Instituto de Astrofisica de Canarias. Funding for SDSS-III has been provided by the Alfred P. Sloan Foundation, the participating institutions, the National Science Foundation, and the U.S. Department of Energy Office of Science. The SDSS-III web site is http://www.sdss3.org/. SDSS-III is managed by the Astrophysical Research Consortium for the Participating Institutions of the SDSS-III Collaboration including the University of Arizona, the Brazilian Participation Group, Brookhaven National Laboratory, University of Cambridge, Carnegie Mellon University, University of Florida, the French Participation Group, the German Participation Group, Harvard University, the Instituto de Astrofisica de Canarias, the Michigan State/Notre Dame/JINA Participation Group, Johns Hopkins University, Lawrence Berkeley National Laboratory, Max Planck Institute for Astrophysics, Max Planck Institute for Extraterrestrial Physics, New Mexico State University, New York University, Ohio State University, Pennsylvania State University, University of Portsmouth, Princeton University, the Spanish Participation Group, University of Tokyo, University of Utah, Vanderbilt University, University of Virginia, University of Washington, and Yale University.

\end{acknowledgements}

\begin{figure*}
\begin{center}
        \includegraphics[width=18.5cm]{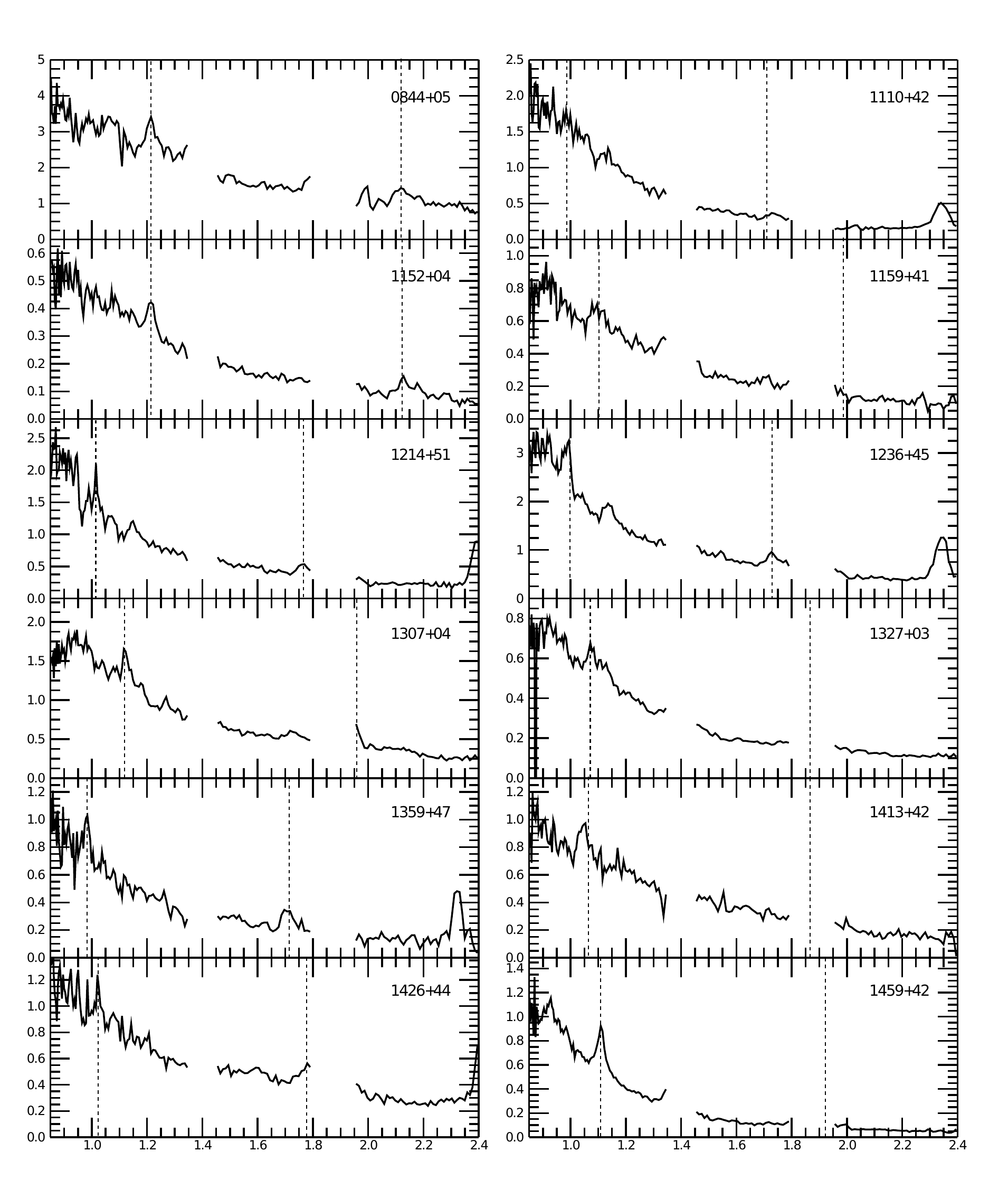}
        \caption{TNG spectra of the RL (first 16 panels) and RQ (second 18 panels) BAL QSOs from our optically bright sample. Units are 10$^{-16}$ erg/s/cm$^{2}$/$\AA$ vs $\mu$m. Regions where atmospheric transmittance becomes critical have been blanked (1.35 - 1.44 $\mu$m and 1.81 - 1.94 $\mu$m). Dashed lines indicate the position of the \mgii~(left side) and \hb~(right side) emission lines at rest-frame.}
\label{spectra}
\end{center}
\end{figure*}
\addtocounter{figure}{-1}
\begin{figure*}
\begin{center}
        \includegraphics[width=18.5cm]{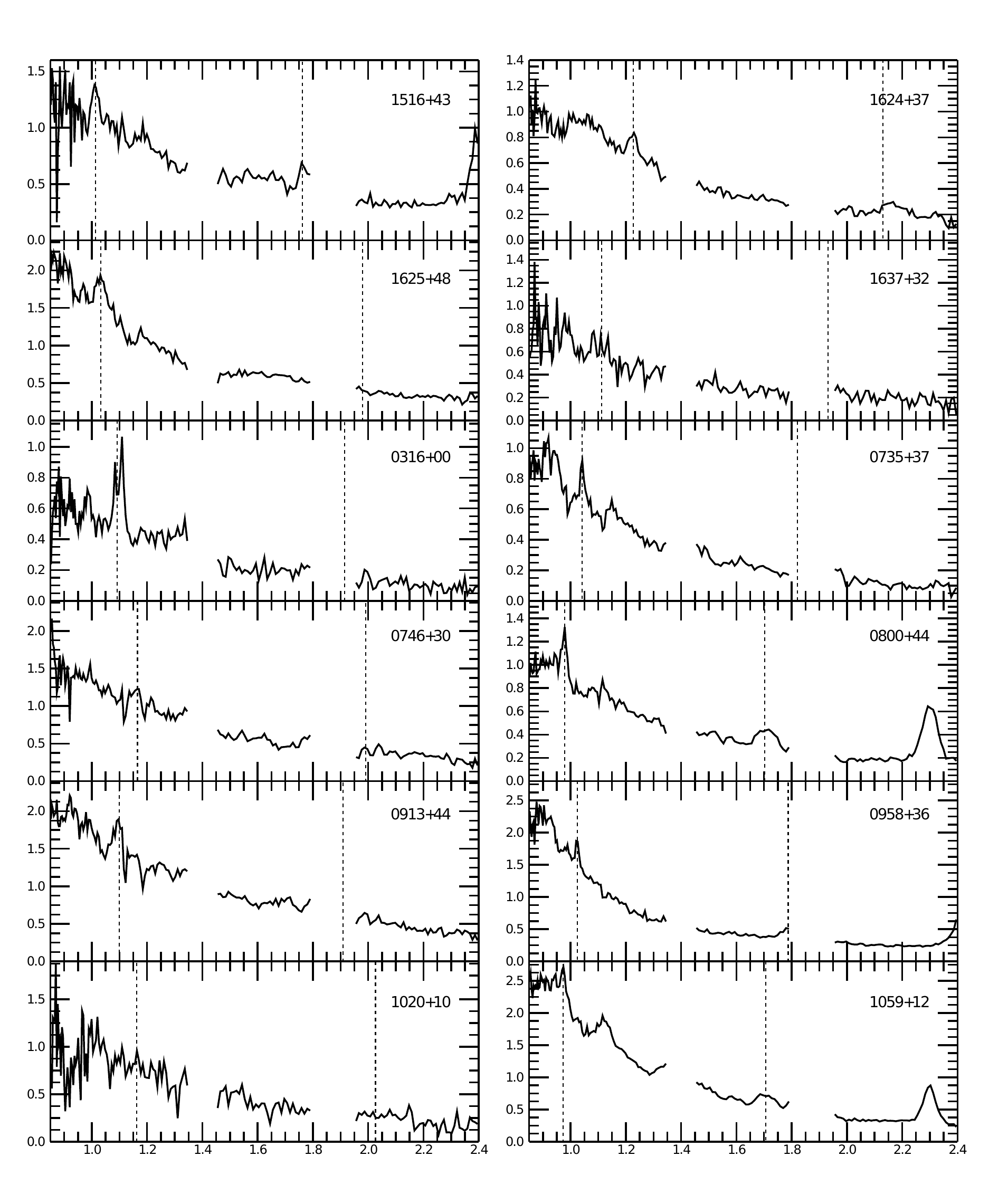}
        \caption{Continued.}
\label{spectra}
\end{center}
\end{figure*}
\addtocounter{figure}{-1}
\begin{figure*}
\begin{center}
        \includegraphics[width=18.3cm]{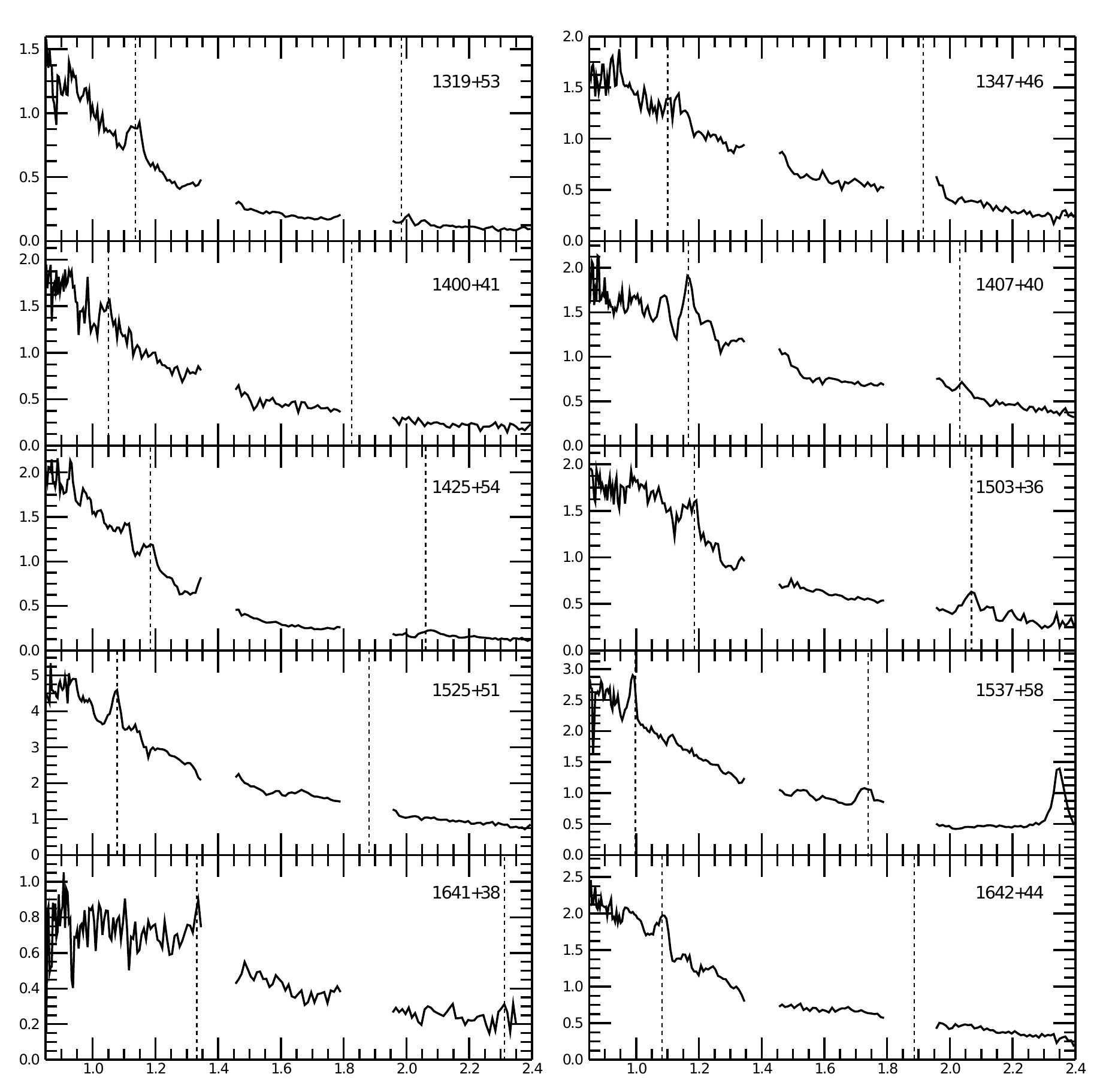}
        \caption{Continued.}
\label{spectra}
\end{center}
\end{figure*}
%
%
%
%

\begin{thebibliography}{}
\bibitem[\protect\citeauthoryear{Adelman-McCarthy et al.}{2006}]{Adelman06} Adelman-McCarthy, J., Agueros, M.A., Allam, S.S. et al. 2006, ApJS, 162, 38
\bibitem[\protect\citeauthoryear{Ahn et al.}{2012}]{Ahn} Ahn, C.P., Alexandroff, R., Allende Prieto, C. et al. 2012, ApJS, 203, 21
\bibitem[\protect\citeauthoryear{Allen et al.}{2011}]{Allen} Allen, J.T., Hewett, P.C., Maddox, N. et al. 2011, MNRAS, 410, 860 
\bibitem[\protect\citeauthoryear{Baffa et al.}{2001}]{Baffa} Baffa C., Comoretto G., Gennari S. et al. 2001, A\&A, 378, 722
\bibitem[\protect\citeauthoryear{Becker et al.}{2000}]{Becker1} Becker, R.H., White, R.L., Gregg, M.D. et al. 2000, ApJ, 538, 72
\bibitem[\protect\citeauthoryear{Becker et al.}{2001}]{Becker2} Becker, R.H., White, R.L., Gregg, M.D. et al. 2001, ApJS, 135, 227
\bibitem[\protect\citeauthoryear{Bentz et al.}{2006}]{Bentz} Bentz, M.C., Peterson, B.M., Pogge et al. 2006, ApJ, 644, 133
\bibitem[\protect\citeauthoryear{Boroson et al.}{2002}]{Boroson} Boroson, T.A. 2002, ApJ, 565, 78
\bibitem[\protect\citeauthoryear{Boroson \& Meyers}{1992}]{Boroson2} Boroson, T., Meyers, K.A. 1992, ApJ, 397, 442
\bibitem[\protect\citeauthoryear{Briggs et al.}{1984}]{Briggs} Briggs, F. H., Turnsheck, D. A. \& Wolfe M. 1984, ApJ, 287, 549 
\bibitem[\protect\citeauthoryear{Bruni et al.}{2012}]{Bruni1} Bruni, G., Mack, K.-H., Salerno, E. et al. 2012, \aap, 542, A13
\bibitem[\protect\citeauthoryear{Bruni et al.}{2013}]{Bruni2} Bruni, G., Dallacasa, D., Mack, K.-H. et al. 2013, A\&A, 554, A94
\bibitem[\protect\citeauthoryear{Cao Orjales et al.}{2012}]{Cao} Cao Orjales, J. M., Stevens, J. A., Jarvis, M. J. et al. 2012, MNRAS, 427, 1209
\bibitem[\protect\citeauthoryear{Crenshaw, Kraemer \& George}{2002}]{Crenshaw} Crenshaw, D.M., Kraemer, S.B. \& George, I. M. 2002, “Mass outflow in AGN: new perspectives”, ASP conference Series, vol. 255
\bibitem[\protect\citeauthoryear{de Kool}{1997}]{deKool} de Kool, M. 1997, in “Mass Ejection from Active Galactic Nuclei”, ASP Conference Series, vol. 128, p. 233
\bibitem[\protect\citeauthoryear{DiPompeo et al.}{2011}]{DiPompeo1} DiPompeo, M. A., Brotherton, M. S., De Breuck, C., et al. 2011, ApJ, 743, 71
\bibitem[\protect\citeauthoryear{DiPompeo et al.}{2013}]{DiPompeo2} DiPompeo, M. A., Runnoe, J. C., Brotherton, M. S. et al. 2013, ApJ 762, 111
\bibitem[\protect\citeauthoryear{Elvis}{2000}]{Elvis} Elvis, M. 2000, ApJ, 545, 63
\bibitem[\protect\citeauthoryear{Filiz Ak et al.}{2012}]{Filiz} Filiz Ak, N., Brandt, W. N., Hall, P. B. et al. 2012, ApJ, 757, 114
\bibitem[\protect\citeauthoryear{Gibson et al.}{2009}]{Gibson} Gibson, R. R., Jiang, L., Brandt, W. N. et al. 2009, ApJ, 692, 758
\bibitem[\protect\citeauthoryear{Gregg at al.}{1996}]{Gregg96} Gregg, M.D., Becker, R.H., White, R.L. et al. 1996, AJ, 112, 2, 407
\bibitem[\protect\citeauthoryear{Gregg at al.}{2006}]{Gregg} Gregg, M.D., Becker, R.H., and de Vries, W. 2006, ApJ, 641, 210
\bibitem[\protect\citeauthoryear{Hall et al.}{2002}]{Hall} Hall, P. B., Anderson, S. F., Strauss, M. A. et al. 2002, ApJS, 141, 267
\bibitem[\protect\citeauthoryear{Hewett \& Foltz}{2003}]{Hewett} Hewett, P.C. \& Foltz, C.B. 2003, AJ, 125, 1784
\bibitem[\protect\citeauthoryear{Kaspi et al.}{2000}]{Kaspi} Kaspi S., Smith P.S., Netzer H., Maoz D., Jannuzi B.T., Giveon U. 2000, ApJ, 533, 631 
\bibitem[\protect\citeauthoryear{Kaspi et al.}{2005}]{Kaspi05} Kaspi, S., Maoz, D., Netzer, H., Peterson, B. M., Vestergaard, M., \& Jannuzi, B. T. 2005, ApJ, 629, 61
\bibitem[\protect\citeauthoryear{Korista et al.}{1993}]{kor93} Korista K.T., Voit G.M., Morris S.L.,  Weymann R.J. 1993, ApJS, 88, 357
\bibitem[\protect\citeauthoryear{Montenegro-Montes et al.}{2008}]{Montenegro} Montenegro-Montes, F.M., Mack, K.-H., Vigotti, M. et al. 2008, MNRAS, 388, 1853
\bibitem[\protect\citeauthoryear{Murray et al.}{1995}]{Murray}Murray, N., Chiang, J., Grossman, S.A., Voit, G.M. 1995, ApJ, 451, 498
\bibitem[\protect\citeauthoryear{O'Dea}{1998}]{ODea} O'Dea, C.P. 1998, PASP, 110, 493
\bibitem[\protect\citeauthoryear{Oliva}{2003}]{Oliva} Oliva E. 2003, Mem Soc. Astron. It., 74, 118
\bibitem[\protect\citeauthoryear{Peterson et al.}{2004}]{Peterson04} Peterson, B.M. 2004, ApJ, 613, 682
\bibitem[\protect\citeauthoryear{Peterson}{2011}]{Peterson11} Peterson, B.M. 2011, PoS(NLS1)032
\bibitem[\protect\citeauthoryear{Proga et al.}{1998}]{Proga1} Proga, D., Stone, J. M., Drew, J. E. 1998, MNRAS, 295, 595
\bibitem[\protect\citeauthoryear{Proga et al.}{2004}]{Proga2} Proga, D., Kallman, T. R. 2004, ApJ, 616, 688
\bibitem[\protect\citeauthoryear{Runnoe et al.}{2013}]{Runnoe} Runnoe, J. C., Ganguly, R., Brotherton, M. S. et al. 2013, MNRAS, 433, 1778
\bibitem[\protect\citeauthoryear{Sanders}{2002}]{Sanders} Sanders, D. B. 2002, ASP Conf. Ser., 284, 411
\bibitem[\protect\citeauthoryear{Schneider et al.}{2010}]{Schneider} Schneider, D. P., Richards, G. T., Hall P. B. et al. 2010, AJ, 139, 2360
\bibitem[\protect\citeauthoryear{Shen et al.}{2011}]{Shen} Shen, Y., Richards, G.T., Strauss M.A. et al. 2011, ApJ, 194, 45
\bibitem[\protect\citeauthoryear{Skrutskie et al.}{2006}]{2mass} Skrutskie, M.F., Cutri, R.M., Stiening, R. et al. 2006, AJ, 131, 1163
\bibitem[\protect\citeauthoryear{Stocke et al.}{1992}]{Stocke} Stocke, J.T., Morris, S.L., Weymann, R.J. et al. 1992, ApJ, 396, 487
\bibitem[\protect\citeauthoryear{Trump et al.}{2006}]{Trump} Trump, J. R., Hall, P. B., Reichard, T. A. et al. 2006, ApJS, 165, 1
\bibitem[\protect\citeauthoryear{Vestergaard et al.}{2006}]{Vestergaard} Vestergaard M., Peterson B.M. 2006, ApJ, 641, 689
\bibitem[\protect\citeauthoryear{Vestergaard \& Osmer}{2009}]{Vestergaard2} Vestergard, M. \& Osmer, P.S. 2009, ApJ, 699, 800
\bibitem[\protect\citeauthoryear{Weymann et al.}{1991}]{Weymann} Weymann, R.J., Morris, S.L., Foltz, C.B. et al. 1991, ApJ, 373, 23

\end{thebibliography}
\end{document}